\documentclass{iau} 
\usepackage{graphicx}
\usepackage{gensymb}
\usepackage{wrapfig}
\usepackage{amsmath,amssymb}
\usepackage{comment}

\title[Disk water megamasers in Seyfert 2 AGN] %% give here short title %%
{On the low detection efficiency of disk water megamasers in Seyfert 2 AGN}

\author[Alberto Masini \& Andrea Comastri]   %% give here short author list %%
{Alberto Masini$^{1,2}$
 \and Andrea Comastri$^1$}

\affiliation{$^1$INAF-Osservatorio Astronomico di Bologna, via Gobetti 93/3, 40129 Bologna, Italy \\[\affilskip]
$^2$Dipartimento di Fisica e Astronomia, Universit\`a  di Bologna, via Gobetti 93/2, 40129 Bologna, Italy \\email: {\tt alberto.masini4@unibo.it}}

\pubyear{2017}
\volume{336}  %% insert here IAU Symposium No.
\jname{Astrophysical Masers: Unlocking the Mysteries of the Universe}
\editors{A. Tarchi, M.J. Reid \& P. Castangia, eds.}
\begin{document}

\maketitle

\begin{abstract}
Disk megamasers are a unique tool to study active galactic nuclei (AGN) sub-pc environment, and precisely measure some of their fundamental parameters. While the majority of disk megamasers are hosted in heavily obscured (i.e., Seyfert 2, Sy2) AGN, the converse is not true, and disk megamasers are very rarely found even in obscured AGN. The very low detection rate of such systems in Sy2 AGN could be due to the geometry of the maser beaming, which requires a strict edge-on condition. We explore some other fundamental factors which could play a role in a volume-limited survey of disk megamasers in Sy2 galaxies, most importantly the radio luminosity.
\end{abstract}

\firstsection % if your document starts with a section,
\section{Introduction}
Many works found a striking correspondence between the presence of a megamaser disk and Compton-thick absorption in the X-ray band (\cite[e.g., Greenhill et al. 2008, Castangia et al. 2013, Masini et al. 2016]{greenhill08,castangia13,masini16}), implying a tight link between megamaser disks and Sy2 AGN, with the masing disk tracing the toroidal obscuring medium invoked by AGN unification models. However, since the maser beaming angle is quite narrow (15\degree-20\degree) and maser amplification occurs preferentially along the disk equatorial plane, disks are detected when they are almost edge-on. This makes them hard to detect even in Sy2 AGN, with an observed detection efficiency $\lesssim1\%$ (Braatz et al. in prep). Although not corrected for the survey sensitivity, this value is worth to be explored.

\section{Covering factors and Radio luminosity}
\begin{figure}[hb]
%\vspace*{-2.0 cm}
\begin{center}
\includegraphics[width=0.496\textwidth]{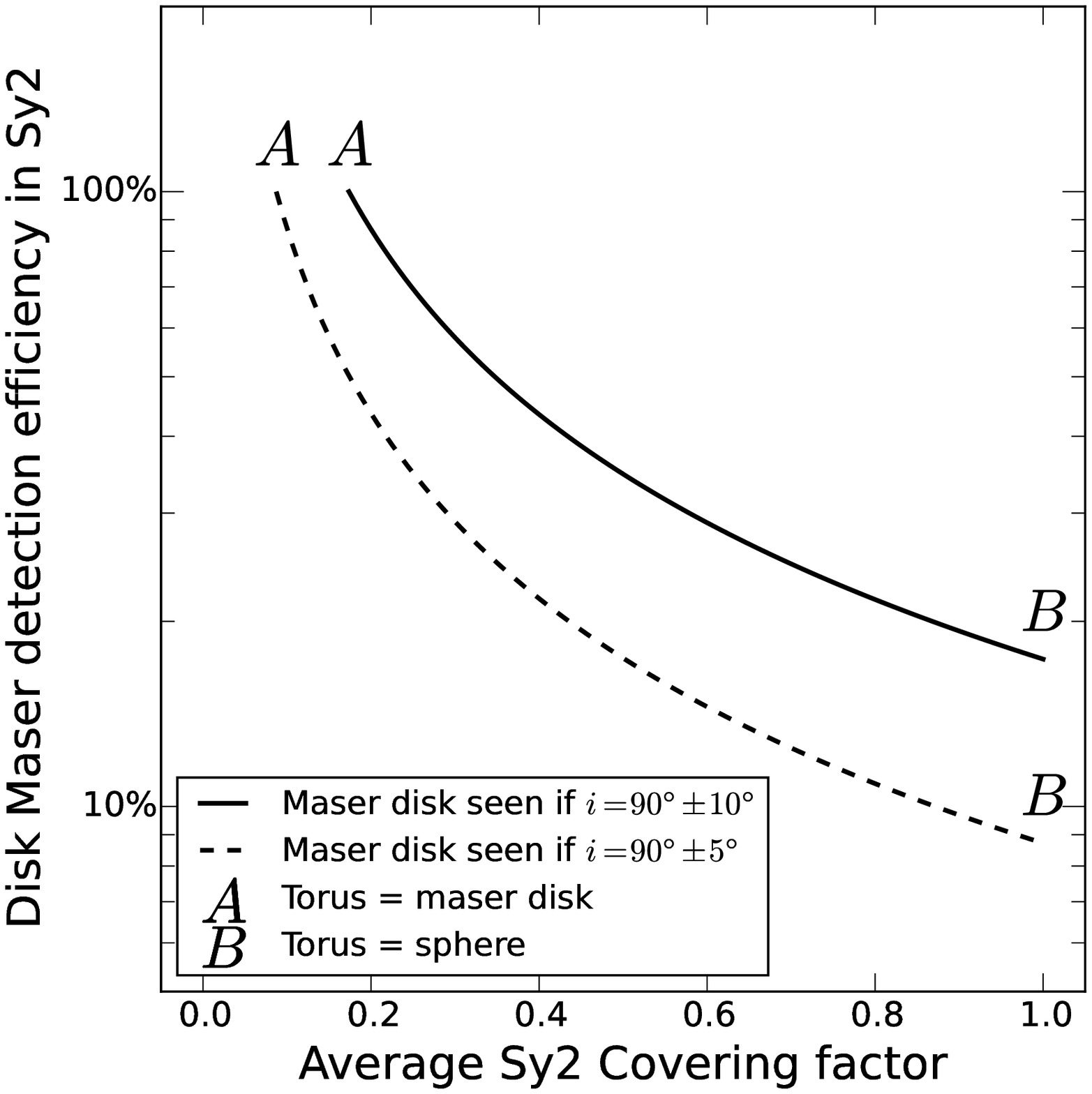} 
\includegraphics[width=0.496\textwidth]{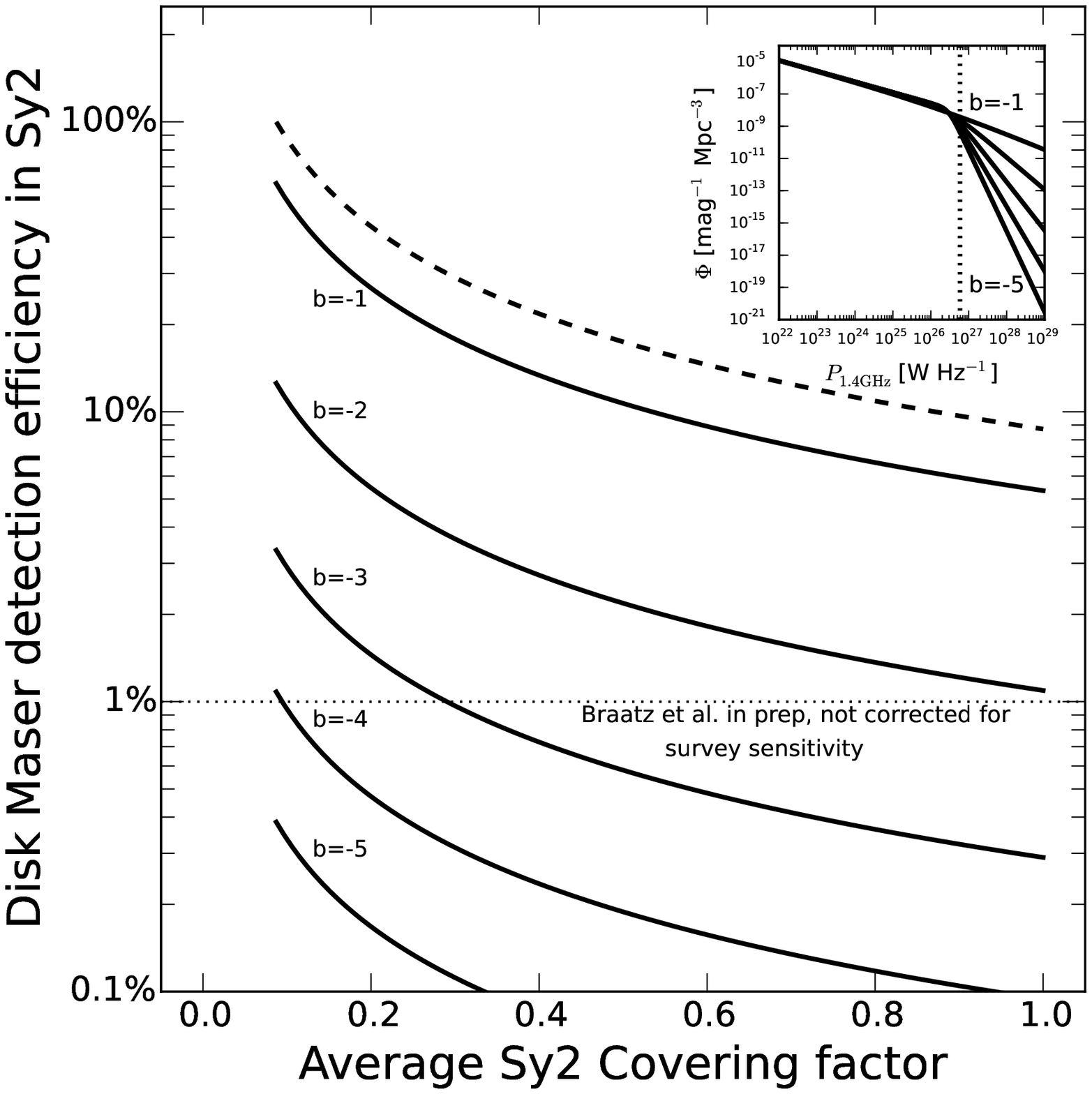}
% \vspace*{-1.0 cm}
 \caption{\textit{Left.} Disk maser detection efficiency as a function of average Sy2 covering factor, assuming two different thickness for the maser disk. \textit{Right.} Effect of different slopes $b$ of the 1.4 GHz LF (inset) on the intrinsic detection efficiency.}
   \label{fig1}
\end{center}
\end{figure}

Assuming that a maser disk is detected if the line of sight angle ranges between $(90 \pm 10)\degree$ with respect to the polar axis, its half-opening angle, defined as the angle between the symmetry axis of the system and the edge of the disk, is $\theta_{\rm disk} = 80\degree$. A Sy2 galaxy, on the other hand, is detected when the torus intercepts the line of sight, and we can analogously define its opening angle $\theta_{\rm tor}$. The average Sy2 covering factor (CF) can range between that of the maser disk itself ($\theta_{\rm tor} = \theta_{\rm disk}$) and a $\textrm{CF}=1$ (i.e. a sphere, $\theta_{\rm tor}=0$), where the covering factor of a toroidal structure with an half-opening angle $\theta$ is defined as $\textrm{CF}=\sin(\pi/2-\theta)$.
The probability of detecting a maser disk in a Sy2 AGN, $P$, is then given by the ratio of the maser disk covering factor with respect to the torus one:
\begin{equation}
P=\frac{\textrm{CF}_{\rm disk}}{\textrm{CF}_{\rm tor}}=\frac{\sin(\pi/2-\theta_{\rm disk})}{\sin(\pi/2-\theta_{\rm tor})}.
\end{equation}
This probability gives the intrinsic detection efficiency (i.e., when correcting for survey completeness) of disk masers in Sy2 AGN. In the left panel of Figure \ref{fig1}, the maser disk detection efficiency is shown as a function of the average Sy2 covering factor (making $\theta_{\rm tor}$ vary, assuming $\theta_{\rm disk} = 80\degree$ and $\theta_{\rm disk} = 85\degree$; black solid and dashed lines, respectively). With this simple geometric argument, we get that the \textit{intrinsic} fraction of maser disks in Sy2 AGN is always higher than $\sim10\%$. This means that, in a volume-limited survey, the intrinsic fraction of Sy2 AGN hosting a disk maser should be $>10\%-20\%$, depending on the effective beaming angle of the maser disk, in contrast with observations. \newline An hard X-ray luminosity threshold $L_{\rm X}$ might be a crucial parameter in order to explain the observed low detection efficiency, and a proper hard X-ray selection may boost the observed detection efficiency in radio surveys to $20\%-30\%$ (see F. Panessa, this volume). \newline The radio power may be another crucial factor to be considered, since there is evidence that megamaser galaxies are a factor of $2-3$ radio brighter than non-megamaser galaxies \cite[(Liu et al., 2017; Zhang et al., 2017)]{liu17,zhang17}. In particular, adopting the radio luminosity function (LF) at 1.4 GHz of Pracy et al. (2016), one can compute the probability for a Sy2 galaxy to have a radio power higher than, e.g., $2P_*$. Depending on the unconstrained high-end of the LF, different probabilities are computed, which lower the intrinsic fraction of maser disks in Sy2 or, equivalently, boost the detection efficiency of radio surveys (right panel of Figure \ref{fig1}). Measuring the exact value of the intrinsic detection efficiency of disk megamasers and the shape of the high frequency radio LF will shed new light on the average Sy2 covering factor.
 
\end{document}